\documentclass[preprints,article,accept,moreauthors,pdftex]{mdpi}

\usepackage{upgreek}
\usepackage{textcomp}
\usepackage{booktabs}
\usepackage{amssymb,amsmath,amsthm}
\usepackage{graphicx}
\usepackage{adjustbox,lipsum} 
\usepackage{esdiff}
\usepackage{physics}
\usepackage[version=3]{mhchem}

\usepackage{siunitx}
\newcommand{\pder}[2][]{\frac{\partial#1}{\partial#2}}

\let\oldsqrt\sqrt
\def\sqrt{\mathpalette\DHLhksqrt}
\def\DHLhksqrt#1#2{%
	\setbox0=\hbox{$#1\oldsqrt{#2\,}$}\dimen0=\ht0
	\advance\dimen0-0.2\ht0
	\setbox2=\hbox{\vrule height\ht0 depth -\dimen0}%
	{\box0\lower0.4pt\box2}}
\makeatletter

\usepackage{array}
\newcolumntype{P}[1]{>{\centering\arraybackslash}p{#1}}

\usepackage{adjustbox}

\usepackage{blindtext}
\usepackage{scrextend}
\addtokomafont{labelinglabel}{\sffamily}

\newsavebox\myboxA
\newsavebox\myboxB
\newlength\mylenA
\newcommand*\xoverline[2][0.75]{%
	\sbox{\myboxA}{$\m@th#2$}%
	\setbox\myboxB\null
	\ht\myboxB=\ht\myboxA%
	\dp\myboxB=\dp\myboxA%
	\wd\myboxB=#1\wd\myboxA
	\sbox\myboxB{$\m@th\overline{\copy\myboxB}$}
	\setlength\mylenA{\the\wd\myboxA}
	\addtolength\mylenA{-\the\wd\myboxB}%
	\ifdim\wd\myboxB<\wd\myboxA%
	\rlap{\hskip 0.5\mylenA\usebox\myboxB}{\usebox\myboxA}%
	\else
	\hskip -0.5\mylenA\rlap{\usebox\myboxA}{\hskip 0.5\mylenA\usebox\myboxB}%
	\fi}
\makeatother

\firstpage{1} 
\pubvolume{xx}
\issuenum{1}
\articlenumber{5}
\pubyear{2019}
\copyrightyear{2019}
\history{Received: date; Accepted: date; Published: date}
\Title{Mathematical Modelling of Hydrophilic Ionic Fertiliser Diffusion in Plant Cuticles: Lipophilic~Surfactant Effects} 

\Author{   {Eloise~C.~Tredenick}\,$^{1,}$*\orcidA{},    {Troy~W.~Farrell}\,   {$^{1,2,}$}*\orcidB{} and W.~Alison~Forster\,   {$^{3}$}}
\AuthorNames{Eloise~C.~Tredenick, Troy~W.~Farrell and W.~Alison~Forster}
\address{%
$^{1}$ \quad School of Mathematical Sciences, Queensland University of Technology, GPO Box 2434, Brisbane,    {Queensland}, 4001, Australia\\
$^{2}$ \quad ARC Centre of Excellence for Mathematical and Statistical Frontiers (ACEMS), \mbox{Queensland University of~Technology}, Brisbane,    {Queensland}, Australia\\
$^{3}$ \quad Plant Protection Chemistry NZ Ltd., PO Box 6282,    {Rotorua}, Bay of Plenty, 3043, New Zealand }
\corres{   {Correspondence: eloise.tredenick@qut.edu.au} (E.C.T.); t.farrell@qut.edu.au (T.W.F.)}

\abstract{The global agricultural industry requires improved efficacy of sprays being applied to weeds and crops to increase financial returns and reduce environmental impact. Enhancing foliar penetration is one way to improve efficacy. Within the plant leaf, the cuticle is the most significant barrier to agrochemical diffusion. It has been noted that a comprehensive set of mechanisms for ionic active ingredient (AI) penetration through plant leaves with surfactants is not well defined, and oils that enhance penetration have been given little attention. The importance of a mechanistic mathematical model has been noted previously in the literature. Two mechanistic mathematical models have been previously developed by the authors, focusing on plant cuticle penetration of calcium chloride through tomato fruit cuticles. The models included ion binding and evaporation with hygroscopic water absorption, along with the ability to vary the AI concentration and type, relative humidity, and plant species. Here, we further develop these models to include lipophilic adjuvant effects, as well as the adsorption and desorption, of compounds on the cuticle surface with a novel Adaptive Competitive Langmuir model. These modifications to a penetration model provide a novel addition to the literature. We validate our theoretical model results against appropriate experimental data, discuss key sensitivities, and relate theoretical predictions to physical mechanisms. The results indicate the addition of the desorption mechanism may be one way to predict increased penetration at late times, and the sensitivity of model parameters compares well to those present in the literature.}

\keyword{plant cuticle; hydrophilic ionic active ingredient; porous diffusion; adsorption; desorption; lipophilic; mathematical model; aqueous pores; surfactant; competitive Langmuir; ion transport}

\begin{document}

\section{Introduction}\label{sec:lit}
The global agricultural industry requires improved efficacy of sprays applied to crops and weeds~\citep{Shaner2014}. While spray application of agrochemicals is known to be effective, it is often inefficient \citep{Knoche1994}. Enhancing the efficacy of agrochemicals has many benefits \citep{Balneaves1993,Schonherr2006,Mckenna2013}. For all foliar-applied agrochemicals, inefficiencies arise during deposition (sprays are not reaching the target) and retention (sprays are reaching the target but are not retained). In the case of systemic pesticides and foliar fertilisers, additional inefficiencies arise during penetration (not all active ingredient (AI) penetrates the plant) and translocation (not all AI within the target plant is transported to the site of biological activity).

The plant cuticle is considered the rate-limiting barrier to foliar penetration of agrochemicals \citep{Schonherr1989}. Aqueous pores are dynamic nanopores within the cuticle that form only in the presence of water \citep{Schonherr2006,Riederer2001,Mays2007new}. Hydrophilic ionic agrochemicals (ionic AIs) and water penetrate the plant cuticle through aqueous pores via diffusion \citep{Baur1999diffusion,Schonherr2006,Schreiber2006}. Ionic AI penetration has major practical importance to the agricultural industry \citep{Schreiber2005}. 

The mechanisms governing penetration of ionic AIs through plant cuticles has been reviewed elsewhere \citep{Tredenick2017,Tredenick2018,Fernandez2017physico,Schonherr2006, Schreiber2005}. Significant factors include plant species variations, ion binding to the cuticle surface, relative humidity, droplet evaporation, and point of deliquescence with hygroscopic water absorption \cite{jeffree20082,Schreiber2006Agcl,Santier1992,Yamada1964,Tang1997,Oxy2014,Hunsche2012}.

The focus of this paper was to further analyse the mechanisms involved in lipophilic surfactant addition to a hydrophilic ionic AI spray solution that penetrates through isolated astomatous plant cuticles. In \citet{Tredenick2018}, when considering the experimental data \citep{Kraemer20092} for penetration of ionic \ce{CaCl2} formulated with the surfactant rapeseed oil (RSO~5), the model was able to predict the overall trends in the experimental data. However, the model was unable to replicate the data trends, where penetration increased at late times, between 24 and 48 h.

Rapeseed oil is an ethoxylated natural oil that exhibits non-ionic surfactant properties and is a versatile emulsifier \citep{Aqnique2016}. The hydrophilic--lipophilic balance (HLB) of RSO~5 is 5.2 \citep{Kraemer20092}, indicating that it is lipophilic, or water-insoluble. A comprehensive set of transport mechanisms for ionic AI penetration through plant leaves, in the presence of surfactants, is not well defined \citep{Hess2000, Wang2007, stock1993possible}. The~mode of action of oils enhancing ionic AI penetration through plant leaves has been given little attention \citep{Wang2007}. The concentration of penetrated adjuvant is rarely measured alongside ionic AI penetration in plant studies, RSO penetration has not been measured to date and the transport of RSO through plant leaves is not fully understood. However, some mechanisms have been identified. Surfactants are known to decrease an ionic AI's point of deliquescence \citep{Chen2001}, allowing further penetration to occur at lower relative humidity and can change the contact area, contact angle, and surface tension of applied droplets \citep{Gaskin2005}. Adjuvants, including surfactants, adsorb to various surfaces \citep{Zhang2017Soft,Ahmadi2015,Bilaowas13static,Peirce2016,Zhang2006}. Aqueous surfactants Triton X-100, SDS (sodium dodecylsulfate), and DTAB (dodecyltrimethylammonium bromide) adsorb to different amounts on adaxial and abaxial wheat leaf (\textit{Triticum aestivum}) surfaces \citep{Zhang2017Soft}. These results were confirmed \citep{Zhang2006} on cabbage (\textit{Brasslca oleracea}) and wheat (\textit{Triticum aestivum}) leaves. They note that adsorption of non-ionic surfactants on hydrophobic surfaces, such as plant cuticles, may lead to a complete replacement of the surface by the adsorbed surfactant as a monolayer, affecting the surface mechanisms. We note the possibility that the adsorbed surfactant could play a role in changing the cuticle surface chemistry, potentially influencing the adsorption of ionic AI, and further research is required.

As RSO~5 has a HLB that is lipophilic, we will assume its transport can appropriately be described as a lipophilic compound. Lipophilic non-electrolytes are compounds that can traverse the plant cuticle exclusively via the lipophilic pathway \citep{Hess2000,Schonherr2006,Schreiber2001}. The lipophilic pathway is made of cutin and wax and is governed by different mechanisms to transport in aqueous pores. Penetration of lipophilic compounds through plant cuticles is a three step process: sorption into the cuticular lipids, diffusion across the cuticular membrane and desorption into the apoplast of the epidermal cells \citep{Buchholz2006, Kirkwood1999,Schonherr1999}. The~term sorption is commonly used as this term is nonspecific and does not imply the location or nature of the interactions of the lipophilic compound within the cuticle membrane \citep{Bukovac1993Char}. Sorption~process can be described by an adsorption isotherm \citep{schreiber1992uptake}. Lipophilic compounds diffuse through the cuticle by jumping into voids or defects that arise due to molecular motion by the polymer segments or \mbox{chains \citep{Schreiber2001,Schonherr2006}}. \citet{Schonherr1988} studied the desorption of a lipophilic compound in various isolated cuticles. They found that significant desorption, between 68$\%$ to 90$\%$ of the applied amount, occurs from the inner cuticle surface to the water bath. Desorption also occurs from the outer cuticle surface. We note the mechanism where lipophilic compounds desorb from the cuticle surface will play an important role in developing the surface chemistry within this work.

Lipophilic compound penetration via the lipophilic pathway and hydrophilic ionic compound penetration via the aqueous pathway are governed by distinct mechanisms and each compound uses its pathway exclusively. In addition to the mechanisms discussed earlier, penetration of hydrophilic ionic AIs through plant cuticles is independent of temperature and plasticisers (accelerators), only weakly affected by wax extraction and less molecular size-selective compared to the lipophilic pathway \cite{Schreiber2005,Schonherr2000,Schonherr2001ca}. It has been noted that for lipophilic compounds, when the temperature increases, voids appear and disappear more frequently, which leads to diffusion rates greatly increasing \cite{Schonherr2006,Buchholz2006}. Lipophilic compounds mobility is significantly influenced by lipophilic accelerators, which dissolve in the cutin and wax domains and act as plasticisers \cite{Schreiber2005,Riederer1995,Buchholz2000,Buchholz2006}.

Surfactants have the ability to increase penetration of ionic AIs, especially at late times. \mbox{\citet{Kraemer20092}} measures penetration of ionic \ce{CaCl2}, which has been formulated with the lipophilic surfactant RSO~5, through isolated astomatous tomato fruit (\textit{Solanum lycopersicum} L., cultivar `Panovy') cuticles. If we focus on the penetration at late times, between 24 and 48 h, when \ce{CaCl2} is formulated with RSO~5, penetration increases by 5.7$\%$ more than without RSO~5. This increase in penetration at late times is confirmed \citep{Gauvrit2007}, with glyphosate penetration and ethoxylated RSO, along with other adjuvants with a range of HLBs, through barley (\textit{Hordeum vulgare} L., cv.~Plaisant) and ryegrass (\textit{Lolium multiflorum} Lam.) leaves. They found a similar penetration profile, where penetration of glyphosate significantly increased to 72 h, and in some cases, a $40\%$ increase was obtained between 24 and 72 h. \citet{Coret1993} also confirmed this trend with glyphosate and non-ionic adjuvant penetration through isolated tomato fruit cuticles, where a mean of $20\%$ additional penetration was obtained between 20 and 100 h.

The mathematical models for ionic penetration in plant leaves and cuticles have been reviewed elsewhere \citep{Tredenick2017,Tredenick2018,Forster2004,Trapp2004}. The authors \citet{Tredenick2017} and \citet{Tredenick2018} previously introduced two mechanistic models to simulate ionic penetration in plant cuticles and incorporated pore swelling, relative humidity, droplet evaporation, and hygroscopic water absorption that is influenced by the point of deliquescence.

Mechanistic models for adsorption and desorption, with a moving droplet radius due to evaporation, are limited in the well-established literature. We previously presented a model in \citet{Tredenick2018}, where ions bind to the leaf surface under the moving droplet radius, but do not desorb. Several adsorption--desorption models are available in the well-established literature, such as the competitive Langmuir model \citep{Butler1929,Markham1931} and an adsorption--desorption model for chemicals in \mbox{soil \citep{Cameron1977}}. However, the models in \citep{Butler1929,Markham1931,Cameron1977} do not consider desorption with a moving droplet radius that is undergoing evaporation and \citet{Tredenick2018} does not include desorption. We wish to consider this case for the adsorption and desorption of agrochemicals, which includes the moving droplet radius due to evaporation on the cuticle surface and create a novel mechanistic model.

There are several models on penetration of lipophilic compounds through plant tissues present in the literature. Previous modelling reviews are published elsewhere \citep{Forster2004,Trapp2004}. Some models incorporate diffusion \citep{Keymeulen1995,Trapp2007,Riederer2002,Brazee2004}, while others employ empirical expressions \citep{Schonherr1994}. Whole plant leaf \citep{Keymeulen1995,Trapp2007,Riederer2002} and isolated cuticle models \citep{Brazee2004,Schonherr1994} have been studied. The empirical model \citep{Schonherr1994} assumes equilibrium has been reached and relies on many experimental parameters, such as partitioning coefficients. Penetration~in plant leaves is related to the physiochemical properties of the lipophilic compound, especially molecular size and lipophilicity, but penetration cannot be predicted by either property \citep{Wang2007}. Therefore, a mechanistic model would be an advantage over an empirical model that relies solely on such parameters. 

\citet{Keymeulen1995} employed a model commonly applied in biological applications, based on Fick's first law of diffusion and Trapp and co-workers have modified this formulation to account for whole plant transport in several publications \citep{Trapp2007,Legind2011}. These models rely on many experimental parameters and cannot be applied to the experimental setup considered in \mbox{\citet{Kraemer20092}}. Three~experimental systems have been developed \citep{Riederer2002} to simplify plant penetration. These~experimental systems are a novel way to test the governing mechanisms as they simplify the complex process of plant leaf penetration. The systems were simulated using finite-element techniques with Fick's first law of diffusion with convection of vapour, incorporating wind. However, they focused on leaf and stomatal transport and applied different experimental techniques to those in \mbox{\citet{Kraemer20092}}, so the models have limited applicability here.

We wished to consider the transport of lipophilic adjuvants by creating a novel mechanistic model that accounts for adsorption, desorption, and diffusion mechanisms and allows for any plant species, lipophilic adjuvant and concentration, temperature, and relative humidity to be used. We aimed to simulate the complex governing mechanisms involved in surfactant enhanced hydrophilic ionic AI penetration through isolated astomatous plant cuticles by utilising a predictive mathematical model. We will expand upon the models in previous works \citep{Tredenick2017,Tredenick2018} by including adsorption and desorption of the ionic AI and lipophilic surfactant from the cuticle surface. In this work, we will investigate the mechanism, where the lipophilic adjuvant causes the ionic AI to desorb from the cuticle surface, and this may increase penetration at late times. Without lipophilic adjuvant, desorption of ionic AI is not possible. We are motivated to describe the increased penetration seen in \mbox{the \citet{Kraemer20092}} experimental data at late times, and this was the key objective of this work. The modelling results were validated against experimental data \citep{Kraemer20092} and a sensitivity analysis was conducted.

\section{Model Framework}\label{sec:model3}

The model takes the form of a quasi-one-dimensional, diffusion model. We will briefly describe the modelling formulation, that is based on the authors previous works \citep{Tredenick2017,Tredenick2018} and we refer the reader to \citet{Tredenick2018} for a full description of the auxiliary equations governing evaporation in Appendix \ref{Appendix1}. We account for three components, hydrophilic ionic active ingredient (AI), water (\ce{H2O}), and lipophilic adjuvant (ADJ). The ionic AI can adsorb and desorb from the cuticle surface and the free ionic AI in the droplet solution can diffuse through the aqueous pores. The lipophilic adjuvant can adsorb and desorb from the cuticle surface, then the adsorbed adjuvant can diffuse through the cuticle via the lipophilic pathway. Water can diffuse through both the aqueous pores and the lipophilic pathway \citep{Schreiber2001,Schonherr2006} within the cuticle. The authors previous model in \citet{Tredenick2018} is simplified here by assuming the porosity does not significantly change in time, to focus and expand on the adsorption mechanism. Pore swelling may be influential under other environmental, initial and boundary conditions, as discussed \citet{Tredenick2017} and \citet{Tredenick2018}, and this will be the subject of future works. All three components change primarily along the cuticle membrane thickness, $x$ ($0 \leq x \leq b$). Initially, a droplet with known contact angle, volume, radius, concentration of ionic AI, and adjuvant is placed on the outer cuticle surface, at $x=0$. A well stirred water bath exists at the inner cuticle surface, at $x=b$. The model, including the variables and parameters, as described in Table \ref{Parameters_Pr3}, governing partial differential equations, initial conditions (ICs), boundary conditions (BCs), and parameters, is as follows:
\begin{align}
		\pder[ c_{\text{AI}}]{t} & = D_{\text{AI}} \  \pdv[2]{ c_{\text{AI}}  }{x} , & 0<x<b, \ t>0, \label{ai3}  \\  
		\pder[ c_{\scriptscriptstyle \text{H$_2$O}}  ]{t} & = \left( D_{\scriptscriptstyle \text{H$_2$O,AqP}} \ + \ D_{\scriptscriptstyle \text{H$_2$O,L}} \right) \  \pdv[2]{ c_{\scriptscriptstyle \text{H$_2$O}} }{x} , & 0<x<b, \ t>0, \label{h2o3}  \\
		\pder[ \Gamma^{\text{ads}}_{\text{ADJ}}]{t} & = D_{\text{ADJ}} \  \pdv[2]{ \Gamma^{\text{ads}}_{\text{ADJ}} }{x} , & 0<x<b, \ t>0, \label{ADJ3}\\
	\text{ICs}: \quad  c_{\text{AI}}(x,0) & = 0,   & 0<x \leq b,   \label{IVPai3}\\
	c_{\text{AI}}(0,0) & = c_{\text{AI,0}}^{\text{drop}} ,   \label{IVPaidrop3}\\
		\Gamma^{\text{ads}}_{\text{AI}}(0) & = 0,    \label{IVPAi3}\\
	c_{\scriptscriptstyle \text{H$_2$O}}(x,0) & = c^{\text{pure}}_{\scriptscriptstyle \text{H$_2$O}}, & 0 < x \leq b,  \label{IVPh2o_3}\\
	c_{\scriptscriptstyle \text{H$_2$O}}(0,0) & = \frac{1-\bar{v}_{\text{AI}} \ c_{\text{AI}}(0,0)  -\bar{v}_{\text{ADJ}} \ c_{\text{ADJ}}(0,0) }{\bar{v}_{\scriptscriptstyle \text{H$_2$O}}}          ,  \label{IVPh2o3}\\
	\Gamma^{\text{ads}}_{\text{ADJ}}(x,0) & = 0,   & 0 \leq x \leq b,   \label{IVPADJ3}
	\end{align}
\begin{alignat}{3}
&   \qquad \qquad \qquad \qquad \qquad c_{\text{ADJ}}(0)  && = c_{\text{ADJ,0}}^{\text{drop}} , \label{IVPADJdrop3}\\
&\text{BC - AI (bath)}: \qquad \quad c_{\text{AI}}(b,t) && = 0, && t>0,    \label{BVCaibath3} \\
&\text{BC - H$_2$O (drop)}:  \qquad c_{\scriptscriptstyle \text{H$_2$O}}(0,t) && =  \frac{1-\bar{v}_{\text{AI}} \ c_{\text{AI}}(0,t) -\bar{v}_{\text{ADJ}} \ c_{\text{ADJ}}(0,t)   }{\bar{v}_{\scriptscriptstyle \text{H$_2$O}}}, \quad  &&   t>0,   \label{BVCh2odrop3}\\
&\text{BC - H$_2$O (bath)}:  \qquad  c_{\scriptscriptstyle \text{H$_2$O}}(b,t) && = c^{\text{pure}}_{\scriptscriptstyle \text{H$_2$O}}, && t>0,  \label{BVCh2obath3} \\
&\text{BC - ADJ (bath)}:  \qquad  \Gamma^{\text{ads}}_{\text{ADJ}}(b,t) && = 0, && t>0,    \label{BVCADJbath3} 
\end{alignat}
\begin{align}
\text{Parameters}:     \qquad \varepsilon_{\text{AqP}}   & = \pi \left( \dfrac{ r_{\text{p}}^{\text{max}}  } {L}  \left( \sqrt{n_{0}}+1 \right) \right ) ^{2},   \label{eps_3}\\
D_{\text{AI}} &  = D_{\text{AI}}^{\text{bulk}} \ \varepsilon_{\text{AqP}} \ ^{ \left( \frac{F_{\text{s}}} {2-F_{\text{s}} } \right)}, \label{diffai3}\\
D_{\scriptscriptstyle \text{H$_2$O,L}} & = D_{\scriptscriptstyle \text{H$_2$O}}^{\text{bulk}} \ \varepsilon_{\text{L}} \ ^{ \left( \frac{F_{\text{s}}} {2-F_{\text{s}} } \right)},   \label{diffh2o3}\\
D_{\scriptscriptstyle \text{H$_2$O,AqP}} &  = D_{\scriptscriptstyle \text{H$_2$O}}^{\text{bulk}} \ \varepsilon_{\text{AqP}} \ ^{ \left( \frac{F_{\text{s}}} {2-F_{\text{s}} } \right)},   \label{diffh2o4}\\
D_{\text{ADJ}} &  = D_{\text{ADJ}}^{\text{bulk}} \ \varepsilon_{\text{L}} \ ^{ \left( \frac{F_{\text{s}}} {2-F_{\text{s}} } \right)},  \label{diffADJ3}
\end{align}
\begin{align}
\text{CCR mode}:  \quad \quad    \diff{\theta}{t}  & = - \dfrac { \Lambda \ \left(  1 + \cos(\theta) \right)^2 \ f(\theta)   }{r_{\text{drop,0}}^2}, & 0 < t \leq t_{\text{rec}}, \label{CCR_theta2}\\
\diff{V^{\text{drop}}_{\scriptscriptstyle \text{H$_2$O}}  } {t} &= - \pi \ \Lambda \ r_{\text{drop,0}} \ f(\theta)  ,& 0 < t \leq t_{\text{rec}}. \label{CCR_V2}
\end{align}
\begin{multline}
\text{CCA mode}: \quad \diff{V^{\text{drop}}_{\scriptscriptstyle \text{H$_2$O}}} { t} = - \pi \ \Lambda  \ f(\theta_{\text{rec}}) \left(  \dfrac{3 \ g(\theta_{\text{rec}}) \ V^{\text{drop}}_{\scriptscriptstyle \text{H$_2$O}} }{\pi}  \right) ^{\frac{1}{3}}  \\
\times    \left[ \dfrac{ \xoverline{\chi}} {V_0 }  \ V^{\text{drop}}_{\scriptscriptstyle \text{H$_2$O}} \ \left( \dfrac{     V^{\text{drop}}_{\scriptscriptstyle \text{H$_2$O}}  }{ V_{\text{Del}} } - 1  \right) \right] \ \left( 1 - \dfrac{ c_{\text{AI}}(0,t) }{  c_{\text{POD}} } \right)     \\
 + \left. \dfrac{ \left( \eta_{\text{pore}} \ A_{\Pi} \ \varepsilon_{\text{AqP}} \ D_{\scriptscriptstyle \text{H$_2$O,AqP}}       \ + \ \varepsilon_{\text{L}} \ D_{\scriptscriptstyle \text{H$_2$O,L}}     \right) \ M_{\scriptscriptstyle \text{w,H$_2$O}} \  A_{\text{drop}} }     {\rho_{\scriptscriptstyle \text{H$_2$O}}} \ \pder[{ c_{\scriptscriptstyle \text{H$_2$O}}}]{x} \right\vert_{x=0} , \  t > t_{\text{rec}}, \label{CCA_V1_3}
\end{multline}
\begin{align}
A_{\text{drop}}(t) & =\pi^{\frac{1}{3}} \left(  3 \ g(\theta) \ V^{\text{drop}}_{\scriptscriptstyle \text{H$_2$O}}(t)  \right) ^{\frac{2}{3}},  \label{Area_drop}  \\
m (t) & = -  \left. \ 10^6 \ M_{\text{w,AI}} \  \eta_{\text{pore}} \ A_{\Pi} \ n_{\text{drops}} \ \varepsilon_{\text{AqP}} \ D_{\text{AI}}  \ \int\limits_0^{t_{\text{final}}}  \ A_{\text{drop}} \  \pder[{ c_{\text{AI}}}]{x} \ dt  \right\vert_{x=b}    , \label{eq:penetration3}\\
\% \text{ Ca penetration} (t) &= \dfrac{m(t) \ 100 \%} { 10^6  \ 10^3 \ c^{drop}_{AI,0} \ n_{\text{drops}}  \  V_0} . \label{eq:uptakepercentage3}
\end{align}

\begin{table} [H]
	\vspace{-25pt} 
	\caption[Model parameters]{Model parameters.}	 	\label{Parameters_Pr3}
		\centering
\tablesize{\footnotesize}
	\begin{tabular}{  p{1.3cm}  p{5cm}  p{2.7cm} p{5.8cm} } 
		\toprule

		{ \textbf{Parameter}} & {\textbf{Definition}} & 		{\textbf{Value and Units}} &  {\textbf{Comments}} \\ \midrule

		{  $A_{\Pi}$}        &  {Control volume area }       & 
		{  m$^2$}          &           \\  	
		
		{  $A_{\text{drop}}(t)$}        &  {  Drop surface contact area}       & 
		{  m$^2$}          &  {  Surface contact area of drop on cuticle surface}          \\  
		
		{  $A_{\text{drop,0}}$}        &  { Initial drop surface contact area}       & 
		{  m$^2$}          &  {  Surface contact area of drop on cuticle surface, \citep{Erbil2002}}          \\  
		
		{  AI}        &  { Active ingredient }       & 
		{  }          &  {   }          \\ 
		
		{  $b$}        &  {  Thickness of cuticle}       & 
		{     {$1.87 \times 10^{-5}$}~m}          &  {  \citep{Chamel1991}}          \\ 
		{  $c^{\text{drop}}_{\text{AI,0}}$} & {  Concentration of AI in drop at $t=0$} & 
		{  mol$/$m$^3$} &  {  \citep{Kraemer20092}} \\ 
		
			$c^{\text{drop}}_{\text{ADJ,0}}   $& Concentration of adjuvant in drop at $t = 0$ & 1~mol$/$m$^3$ (1~g$/$L)&\\

		{  $c^{\text{pure}}_{\scriptscriptstyle \text{H$_2$O}}$}  &  { Pure water concentration at $20\,^{\circ}\mathrm{C}$ and $t=0$ }       &
		{  55,409.78~mol$/$m$^3$}          &  {  calculated  }          \\  
		
		{  $c_{\text{POD}}$}        &  {Point of deliquescence concentration}       & 
		{  mol$/$m$^3$}          &  {\citep{Tredenick2018} }          \\ 
		
		{CCR}        &  { Constant contact radius evaporation mode }       & 
		{  }          &  {   }          \\ 	
		
		{CCA}        &  { Constant contact angle evaporation mode }       & 
		{  }          &  {   }          \\

		{  $c_i(x,t)$}        &  {  Concentration of component $i$}       & 
		{  mol$/$m$^3$}          &  {  }          \\  	
		
		{  $D_{\text{AI}}^{\text{bulk}}$} & {  Self/bulk diffusion coefficient of AI} & 
		{     {7.93$ \times 10^{-10}$~m$^2/$s}} &  {For CaCl$_{2}$, Ca$^{2+}$ diffuses the slowest, so Ca$^{2+}$ value is used, \citep{Yuan1974} } \\

		{  $D_{\scriptscriptstyle \text{H$_2$O}}^{\text{bulk}}$}        &  {  Self/bulk diffusion coefficient of water}       & 
		{    { 2.299 $\times 10^{-9}$~m$^2/$s}}          &  {  \citep{Holz2000}}          \\   
		
		$D_{\text{ADJ}}^{\text{bulk}} $ & Self/bulk diffusion coefficient of adjuvant &   {7.93 $\times 10^{-12}$~m$^2/$s} & (fitted)\\
		
		{  $D_{\text{evap}}$}        &  {  Diffusivity of water in air}       & 
		{     {2.4 $\times 10^{-5}$~m$^2/$s} }          &  {  \citep{Semenov2013}}          \\  		
		
		{  $D_i(x,t)$} & {  Diffusivity of component $i$} & 
		{  m$^2/$s} &  {  \citep{Liu2001}} \\ 
		
			{  $F_{\text{s}}$} & {  Fractal scaling dimension} & 	 {    1.15 (-)} &  {  $1< F_{s}<2$ (fitted)} \\ 
		
		{  $f(\theta)$ } & { Functional variation of $\theta$   } & 
		{   } &  {\citep{Popov2005}  } \\  
		
		{  $g(\theta)$ } & { Functional of $\theta$   } & 
		{   } &  {\citep{Popov2005,Dash2013}  } \\

		{  $H$} & {  Relative humidity} & 
		{  0.7 (70\%)} &  {   \citep{Kraemer20092}} \\ 
		
			HLB  & Hydrophilic Lipophilic Balance  &  &    \\ 
		
		{  $i$} & {  Component AI (CaCl$_{2}$), $\text{H$_2$O}$ or ADJ (RSO~5)} & 
		{  } &  {  } \\  
		
		$k_1$ & Adsorption rate constant ionic AI &    {4.2 $\times 10^{-6}$~m$^3/$}(s mol) &(fitted) \\
		$k_2$ & Desorption rate constant ionic AI &    {6.5 $\times 10^{-8}$~m$^3/$}(s mol) & (fitted)\\
		$k_3$ & Adsorption rate constant adjuvant & 2    {$\times 10^{-8}$~m$^3/$}(s mol) &(fitted) \\
		$k_4$ & Desorption rate constant adjuvant &    {5 $\times 10^{-6}$~m$^3/$}(s mol) & (fitted)\\
				$k_5$ & Kinetic rate constant ionic AI &    {5.9 $\times 10^{-10}$~m$^3$} &(fitted) \\
		$k_6$  & Kinetic rate constant adjuvant &    {8 $\times 10^{-9}$~m$^3$} &(fitted) \\
	$\xoverline{k}_5$	 & Kinetic rate constant ionic AI & mol &{calculated}\\
		$\xoverline{k}_6$	& Kinetic rate constant adjuvant & mol &{calculated} \\

		{  $L$}        &  {  Control volume length}       & 
		{  1~m}          &  {  }          \\

		{  $M_{\text{w,AI}}$} & {  Molecular weight of \ce{CaCl2} } & 
		{  110.98~g$/$mol } &  {  } \\ 
		
		{  $M_{\text{w,ADJ}}$} & {  Molecular weight of RSO~5 } & 	{  992~g$/$mol } &  {\, \citep{Anastopoulos2009transesterification}} \\
		
				{  $M_{\scriptscriptstyle \text{w,H$_2$O}}$} & {  Molecular weight H$_2$O} & 
		{  $18.015$~g$/$mol} &  {  } \\ 
		
		{$m_\infty$} & { Equilibrium mass of water absorbed per \ce{CaCl2} applied } & 
		{ g$_{\scriptscriptstyle \text{H$_2$O}}/$g$_{\text{AI}}$ } &  {\citep{Tredenick2018} } \\
		
		{  $N_{\text{A}}$}        &  {  Avogadro constant}       & 
		{   $6.02214 \times 10^{23}$~mol$^{-1}$}          &  { }          \\ 
		
		{  $n_0$}        &  {  Number of aqueous pores on area $L^2$ (1~m$^2$) of cuticle}       & 
		{  
			(-) }          &  { 
			$n_0 = \eta_{\text{pore}} \ L^2$ }          \\  
		
		{  $P_{\text{v}}$}        &  {  Saturated water vapour pressure in air at $20\,^{\circ}\mathrm{C}$}       & 
		{  2338.8~Pa}          &  {\, \citep{Lide2004}}          \\
		
		{   }          &  { $32\%$RH for \ce{CaCl2} \citep{Kolthoff1969} and $27\%$RH for \ce{CaCl2} with RSO~5 }\\

		{  R}        &  {  Gas constant}       & 
		{  8.3145~Pa$\cdot$m$^3/$K$/$mol}          &  { }\\	
		
		{  RSO}        &  {  Rapeseed oil surfactant}       & 
		{  }          &  { }\\
				{  $r_{\text{drop,0}}$}        &  {Initial drop contact radius}       & 
		{ m }          &  { Contact radius of drop on cuticle surface \citep{Erbil2002}}          \\  
		
		{  $r_{\scriptscriptstyle \text{H$_2$O}}$}        &  {Van der Waals radius of a water molecule}       & 
		{    { $1.5 \times 10^{-10}$~m}}          &  {  \citep{Schreiber2006Agcl}}          \\  	
		
		{  $r_{\text{p}}^{\text{max}}$}        &  {  Maximum radius of aqueous pores}       & 
	{     {$2.12 \times 10^{-9}$~m}}          &  {  For tomato fruit cuticle, (\citep{Schreiber2009} p.~87)}          \\  
	{  $RH$} & { Relative humidity in text} & 
	{   } &  {   } \\   
	
	{ $t$}        &  { Time}       & {s}          &  { }          \\  
	
	{  $T$} & {  Temperature} & 	{  293.15~K ($20\,^{\circ}\mathrm{C}$)} &  {  \citep{Kraemer20092}} \\ 
	
	{  $u$} & {Bound integration variable} & 	{  } &  { } \\

					\bottomrule
	\end{tabular} 
\end{table}

\begin{table} [H]\ContinuedFloat

\vspace{-25pt} 
\caption{{\em Cont.}}  
		\centering
\tablesize{\footnotesize}
\begin{tabular}{  p{1.3cm}  p{5cm}  p{2.7cm} p{5.8cm} } 
\toprule

{ \textbf{Parameter}} & {\textbf{Definition}} & 		{\textbf{Value and Units}} &  {\textbf{Comments}} \\ \midrule

{  $V_0$} & {  Volume of droplet at $t=0$} & 
{     {1 $\times 10^{-9}$~m$^3$}} &  {  \citep{Kraemer20092}} \\ 
{  $V^{\text{drop}}_{\scriptscriptstyle \text{H$_2$O}}(t)$} & {  Volume of water in droplet at time $t$} & 
{  } &  { } \\ 

{  $V_{\text{Del}}(t)$}        &  { Deliquescent droplet volume}       & 
{  m$^3$}          &  {}          \\

{  $\bar{v}_{\text{AI}}$}        &  {  Partial molar volume CaCl$_{2}$}       & 
{    {$1.6 \times 10^{-5}$~m$^3/$mol }}          &  {  \citep{Oakes1995}}          \\ 

{  $\bar{v}_{\text{ADJ}}$}        &  {  Partial molar volume adjuvant}       & 	  {    {$4.97 \times 10^{-7}$~m$^3/$}mol }          &  { }          \\ 

{$\bar{v}_{\scriptscriptstyle \text{H$_2$O}}$} & {Partial molar volume water} & 
{   {$1.8047 \times 10^{-5}$~m$^3/$mol}} &  {\citep{Zen1957}} \\

				{ $x$} & { Length} & 	{ m} &   \\

		$ \varepsilon_{\text{AqP}} $ & Aqueous pore pathway porosity  &   &   \\ 
		$ \varepsilon_{\text{L}} $ & Lipophilic pathway porosity  & 0.03 & (fitted) \\
		
		$\Gamma^{\text{ads}}_{\text{AI}}(t)$&  Concentration adsorbed ionic AI per droplet area on cuticle surface & mol$/$m$^2$ &\\

	$\Gamma^{\text{ads}}_{\text{ADJ}}(t)$ & Concentration adsorbed adjuvant per droplet area on cuticle surface  & mol$/$m$^2$ & \\  
$ \Gamma_S $ & Saturated adsorbed molecules per droplet area  &400~mol$/$m$^2$ & (fitted) \\

		{  $\eta_{\text{pore}}$}        &  { Density of aqueous pores in cuticle}       & 	  {	$2.18 \times 10^{15}$~m$^{-2}$}         &  {\citep{Tredenick2018} }          \\

		{  $\Lambda$}        &  { Evaporation constants as a function of relative humidity}       & 
		{  m$^2/$s}          &  { }         \\	
		
		{  $\psi$}        &  {  Saturated water vapour concentration as a function of relative humidity}       & 
		{  g$/$m$^3$}          &  {  \citep{Erbil2012}}          \\		
		
		{  $\Phi $}        &  {Point of deliquescence humidity shifting factor for choosing the humidity in $m_\infty$ and $c_{\text{POD}}$}       & 
		{  }          &  {}          \\ 
		
		{  $\rho_{\scriptscriptstyle \text{H$_2$O}}$}        &  {  Liquid density H$_2$O at $20\,^{\circ}\mathrm{C}$}       & 
		{  $9.98207  \times 10^{5}$~g$/$m$^3$}          &  {  \citep{Lide1990} }          \\ 
		
		{  $\rho_{\scriptscriptstyle \text{AI}} $}        &  {  Liquid density of \ce{CaCl2} at $20\,^{\circ}\mathrm{C}$}       & 
		{  $2.16 \times 10^{6}$~g$/$m$^3$}          &  {\citep{Company2003} }          \\ 
		
		{  $\theta(t)$}        &  {  Contact angle of drop on cuticle surface that changes with time}       & 
		{ rads }          &  {}          \\ 
		
		{  $\theta_0$}        &  {  Contact angle of drop on cuticle surface at \quad $t=0$}       & 
		{ rads }          &  {See Table 1 in \cite{Tredenick2018}   }          \\ 
		
		{  $\theta_{\text{rec}}$}        &  {Receding contact angle of drop on cuticle surface}       & 
		{ rads }          &  { See Table 1 in \cite{Tredenick2018}   }          \\ 
		
						{  $ \chi   $}        &  {Logistic decay evaporation constant}       &   {0.0428 L$^2/$g$^2$}            &  {\citep{Tredenick2018} }          \\ 
		
		{  $ \xoverline{\chi}   $}        &  {Logistic decay evaporation term (a constant) as a function of initial concentration of ionic AI}       & 
		{}          &  { calculated}          \\ 
		
		{  $\xi $}        &  {Point of deliquescence humidity shifting factor to incorporate with the addition of adjuvants}       & 
		{  $5\%$RH}          &  {\citep{Tredenick2018} }   \\       

		\bottomrule

	\end{tabular}
\end{table}
AI can travel through plant cuticle aqueous pores via Fickian diffusion and this is described by Equation (\ref{ai3}). Water diffuses through aqueous pores and the lipophilic pathway (Equation (\ref{h2o3})). Therefore, two diffusion coefficients are required, one for the diffusion of water through aqueous pores, $D_{\scriptscriptstyle \text{H$_2$O,AqP}}$ (Equation (\ref{diffh2o4})) and the lipophilic pathway, $D_{\scriptscriptstyle \text{H$_2$O,L}}$ (Equation (\ref{diffh2o3})). The adsorbed lipophilic adjuvant can diffuse through the cuticle via the lipophilic pathway (Equation (\ref{ADJ3})). We note the adsorbed adjuvant diffuses in terms of $\Gamma^{\text{ads}}_{\text{ADJ}}$ in mol$/$m$^2$, unlike ionic AI and water, which diffuse in terms of a concentration, $c$, in mol$/$m$^3$.

Initially, the lipophilic adjuvant has not been adsorbed to the cuticle surface or through the cuticle (Equation (\ref{IVPADJ3})). An initially applied concentration of adjuvant is present on the cuticle surface (Equation (\ref{IVPAi3})). We note the parameters described in Equation (\ref{eps_3})--(\ref{diffADJ3}) are constants, as the porosities are constant. The diffusion coefficient for water in the lipophilic pathway, $D_{\scriptscriptstyle \text{H$_2$O,L}}$ (Equation (\ref{diffh2o3})), is the same formulation as the diffusion coefficient for water in the aqueous pathway, $D_{\scriptscriptstyle \text{H$_2$O,AqP}}$ (Equation (\ref{diffh2o4})), but is based on the porosity of the lipophilic pathway, $\varepsilon_{\text{L}}$. Following from this, the diffusion coefficient for adjuvant in the lipophilic pathway, $D_{\scriptscriptstyle \text{ADJ}}$ (Equation (\ref{diffADJ3})), is also a similar formulation to Equation (\ref{diffh2o4}). Droplet evaporation is governed by Equation (\ref{CCR_theta2})--(\ref{CCA_V1_3}) in constant contact radius mode (CCR) then constant contact angle mode (CCA), which includes hygroscopic water absorption, and further details are available in \citet{Tredenick2018}. Water travels along the two pathways so the two diffusivities ($D_{\scriptscriptstyle \text{H$_2$O,AqP}}$ and $D_{\scriptscriptstyle \text{H$_2$O,L}}$) are included in the term for evaporation on the third line of Equation (\ref{CCA_V1_3}), which accounts for the possibility of water diffusing towards the droplet from the~cuticle.

\begin{figure} [H]
	\centering
	\includegraphics[height=0.11 \textheight,keepaspectratio=true]{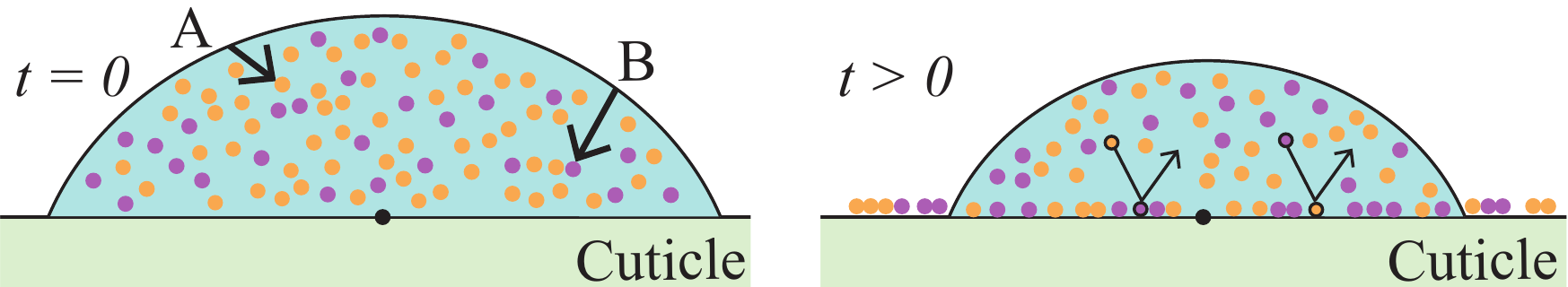}
	\caption[The adsorption and desorption process of ionic AI and adjuvant.]{The Adaptive Competitive Langmuir model that includes adsorption and desorption. The initial condition is at $t=0$ and $t>0$ is a short time later after some droplet evaporation has occurred. Two generic molecules A (orange circles) and B (purple circles) sit on the cuticle surface. Initially, A and B are in a droplet solution. Then as time progresses, A and B adsorb and desorb in a competitive process. As the droplet has moved due to evaporation, some of the molecules have been left behind. We note A in this case is ionic AI and B is lipophilic adjuvant. This figure is a simplification and describes the Adaptive Competitive Langmuir model and moving droplet radius, and we note other processes are involved in droplet evaporation, as described in text (not to scale).}
	\label{ions_released}
\end{figure}

We introduce a mechanism for the droplet boundary condition on the cuticle surface, as depicted in Figure \ref{ions_released}, where ionic AI and lipophilic adjuvant can be adsorbed and desorbed. Focusing on ionic AI, it first adsorbs, then the adsorbed ionic AI can be desorbed by the adjuvant that is adsorbing onto the cuticle surface. More adjuvant adsorbs as the concentration of adjuvant in the droplet increases due to evaporation. The same process occurs for adjuvant and ionic AI. The process is competitive, where ionic AI and adjuvant compete for the same sites on the cuticle surface. The adsorbed ionic AI can only be desorbed by adjuvant in the droplet solution, not ionic AI in solution. The most important feature of this model is that the adjuvant will desorb ionic AI that is adsorbed, and this desorbed ionic AI can then diffuse into the cuticle at later times. The adsorbed ionic AI would otherwise not diffuse through the cuticle.

The ionic AI and adjuvant are contained within a droplet that evaporates and the radius changes with time. This evaporation causes the droplet area to become smaller over time. Only the adsorbed ionic AI or adjuvant that is under the droplet can desorb into the droplet solution. Therefore, a portion of ionic AI and adjuvant that are adsorbed onto the cuticle surface are left behind by the moving droplet radius and are not available for diffusion. As discussed \mbox{in \citet{Tredenick2018,Schonherr2001ca}} and \citet{Schonherr2001potassium}, hygroscopic water absorption in the context of the \citet{Kraemer2009} experimental setup, where relative humidity is $70\%$ and the point of deliquescence of \ce{CaCl2} is $32\%$, will cause the total evaporation and penetration time to increase. If ionic AI is present (as free molecules) in the droplet, the droplet will not completely evaporate. We have assumed that adsorbed ions (that are not free) do not contribute to the hygroscopic water absorption, both inside (adsorbed under the droplet) and outside the droplet (left behind by the moving droplet radius).

The adjuvant adsorbs as a monolayer \citep{Zhang2006}, as does the ionic AI. It is known that adjuvants, including the surfactant RSO~5, adsorb to various surfaces \citep{Zhang2017Soft,Ahmadi2015,Bilaowas13static,Peirce2016,Zhang2006}, lipophilic compounds adsorb and desorb on the cuticle surface \citep{Schonherr1988}, and ionic AIs adsorb to the cuticle surface \citep{Yamada1964}. However, we assume that ionic AIs can desorb and the process is competitive.

We modify the well-known competitive Langmuir formulation \citep{Butler1929,Markham1931}, to describe a model for ionic AI and adjuvant adsorption (ads) and desorption on the cuticle surface. The well-known competitive Langmuir formulation is described elsewhere for comparison (\cite{Tredenick2019Thesis} p.~188). This novel model is known henceforth as the \textit{``Adaptive Competitive Langmuir model"} and adapts to a moving droplet radius that is undergoing evaporation, which influences the amount of compound adsorbed and desorbed.

In Figure \ref{ions_released}, we see the scenario of the Adaptive Competitive Langmuir model. A droplet sits on the cuticle surface that contains ionic AI (A---orange circles) and lipophilic adjuvant (B---purple circles), initially, at $t=0$ and a short time later, at $t>0$. Initially, ionic AI and adjuvant are in a droplet solution. Then after some time, the droplet has partially evaporated and some adsorption of ionic AI and adjuvant has occurred. Ionic AI and adjuvant can desorb each other, competing for the same sites, shown with the arrows. As the droplet has moved due to evaporation, some of the molecules have been left behind. We note that Figure \ref{ions_released} demonstrates the Adaptive Competitive Langmuir model and moving droplet radius, and is an oversimplification, as the ionic AI and adjuvant molecules may not be homogeneously distributed in the droplet solution \citep{Hunsche2012,Zhang2017Soft,Bilaowas13static,Zhang2006}.

Our novel Adaptive Competitive Langmuir model for ionic AI is as follows:
\begin{align}
\diff{\Omega_{\text{AI}}(t)}{t}  & = \ \underbrace{ k_1 \ \overbrace{\left( 1 - \Omega_{\text{AI}}(t)  \right) }^\text{Sites unoccupied by AI} \ c_{\text{AI}}(0,t) }_\text{AI adsorption}    - \underbrace{ k_2  \overbrace{ \Omega_{\text{AI}}(t) }^\text{Sites occupied by AI}   c_{\text{ADJ}}(t)   }_\text{AI desorption} ,\label{eq:comp1}
\end{align}
where $\Omega_{\text{AI}}$ is the fraction of sites occupied by ionic AI on the cuticle surface, $k_1$ is the rate constant for adsorption for ionic AI, $c_{\text{AI}}$ is the concentration of ionic AI in the drop, $k_2$ is the rate constant for desorption of AI and $c_{\text{ADJ}}$ is the concentration of adjuvant in the drop. The term $( 1 - \Omega_{\text{AI}} )$ is the fraction of sites that are either vacant or unoccupied by ionic AI on the cuticle surface. Therefore the first term in Equation (\ref{eq:comp1}) describes the rate of adsorption, based on the sites available (vacant or no ionic AI) and the concentration of ionic AI in the drop, $c_{\text{AI}}$. The second term in Equation (\ref{eq:comp1}) describes the rate of desorption, based on the fraction of sites occupied by ionic AI on the cuticle surface, $\Omega_{\text{AI}}$, and the concentration of adjuvant on the cuticle surface, $c_{\text{ADJ}}$. The ionic AI desorption term of Equation (\ref{eq:comp1}) only allows adsorbed ionic AI to be desorbed by adjuvant in the droplet ($c_{\text{ADJ}}$), not ionic AI in the droplet. 

We then define the fraction of sites occupied by ionic AI under the droplet on the cuticle surface, $\Omega_{\text{AI}}$, which evolves in time with the changing droplet area  during evaporation, as:
\begin{equation}
\Omega_{\text{AI}}(t) = \dfrac{  A_{\text{drop,0}} \ \Gamma^{\text{ads}}_{\text{AI}}(t)} {\Gamma_S \ A_{\text{drop}}(t)},\label{eq:comp2}
\end{equation}
where $A_{\text{drop,0}}$ is the initial droplet contact area, $\Gamma^{\text{ads}}_{\text{AI}}$ is the concentration of adsorbed ionic AI per unit area of cuticle surface under the droplet, $\Gamma^{\text{ads}}_{\text{ADJ}}$ is the concentration of adsorbed adjuvant per unit area of cuticle surface under the droplet, $\Gamma_S$ is the saturated adsorbed molecules per droplet area and $A_{\text{drop}}$ is the droplet area on the cuticle surface.

The equation describing the concentration of adsorbed ionic AI per unit area of cuticle surface under the droplet, $\Gamma^{\text{ads}}_{\text{AI}}$, is found by substituting Equation (\ref{eq:comp2}) into Equation (\ref{eq:comp1}) and simplifying, producing Equation (\ref{eq:AIads}) and an analogous equation for $\Gamma^{\text{ads}}_{\text{ADJ}}$ in Equation (\ref{eq:ADJads}).
\begin{multline}
\diff{  \Gamma^{\text{ads}}_{\text{AI}}(t) }{t}    = \underbrace{ k_1 \ c_{\text{AI}}(0,t) \ \left[ \dfrac{\Gamma_S} {  A_{\text{drop,0}} }   \ A_{\text{drop}}(t) - \Gamma^{\text{ads}}_{\text{AI}}(t)  \right] }_\text{AI adsorption} \\
- \ \underbrace{k_2 \ c_{\text{ADJ}}(t)  \ \Gamma^{\text{ads}}_{\text{AI}}(t) }_\text{AI desorption} + \ \underbrace{  \dfrac{ \Gamma^{\text{ads}}_{\text{AI}}(t)} {  A_{\text{drop}}(t) }  \ \diff{ A_{\text{drop}}(t) }{t},}_\text{Change in droplet area}      \label{eq:AIads}
\end{multline}
\begin{multline}
\diff{  \Gamma^{\text{ads}}_{\text{ADJ}}(0,t) }{t}  = \underbrace{ k_3 \ c_{\text{ADJ}}(t) \ \left[ \dfrac{\Gamma_S} {  A_{\text{drop,0}} }   \ A_{\text{drop}}(t) - \Gamma^{\text{ads}}_{\text{ADJ}}(0,t)   \right] }_\text{ADJ adsorption}\\
- \ \underbrace{ k_4 \ c_{\text{AI}}(0,t)  \ \Gamma^{\text{ads}}_{\text{ADJ}}(0,t) }_\text{ADJ desorption} + \ \underbrace{  \dfrac{ \Gamma^{\text{ads}}_{\text{ADJ}}(0,t)} {  A_{\text{drop}}(t) }  \ \diff{ A_{\text{drop}}(t) }{t} }_\text{Change in droplet area}    \\
+ \underbrace{ \left.    \dfrac{ \varepsilon_{\text{L}} \ A_{\text{drop}}(t)  \ D_{\text{ADJ}}  }{V^{\text{drop}}_{\scriptscriptstyle \text{H$_2$O}}(t)} \ \pder[{  \Gamma^{\text{ads}}_{\text{ADJ}}(x,t)  }]{x} \right\vert_{x=0}  }_\text{Spreads into cuticle}    , \label{eq:ADJads}
\end{multline}
where $k_3$ is the adsorption rate constant for the adjuvant and $k_4$ is the desorption rate constant for the adjuvant. The first term in Equation (\ref{eq:AIads}) accounts for the adsorption of the ionic AI to the cuticle surface based on the concentration of ionic AI in the droplet, $c_{\text{AI}}$, and the sites that are unoccupied by ionic AI under the droplet, $(\Gamma_S A_{\text{drop}} /  A_{\text{drop,0}}  - \Gamma^{\text{ads}}_{\text{AI}})$. The second term accounts for the desorption of ionic AI from the cuticle surface based on the concentration of adjuvant on the droplet surface, $c_{\text{ADJ}}$, and the amount of ionic AI adsorbed per area, $\Gamma^{\text{ads}}_{\text{AI}}$. The third term accounts for the changing droplet area, $A_{\text{drop}}$. Equation (\ref{eq:ADJads}) is similar to Equation (\ref{eq:AIads}), except the last term is added to account for the adsorbed adjuvant, $\Gamma^{\text{ads}}_{\text{ADJ}}$, being transported away from the cuticle surface and into the lipophilic pathway of the cuticle, based on the porosity of the lipophilic pathway, $\varepsilon_{\text{L}}$, the changing area of the droplet, $ A_{\text{drop}}$, the diffusion coefficient for the adjuvant, $D_{\text{ADJ}}$, and the flux of the adsorbed adjuvant at the droplet boundary.

The equations for the concentration of ionic AI and adjuvant in the drop, $c_{\text{AI}}$ and $c_{\text{ADJ}}$, are \mbox{as follows}:
\begin{multline}
\diff{}{t}  \left( V^{\text{drop}}_{\scriptscriptstyle \text{H$_2$O}}(t) \  c_{\text{AI}}(0,t)  \right)  =  \dfrac {\xoverline{k}_5 \ A_{\text{drop,0}}}  {\Gamma_S \ A_{\text{drop}}(t)   }   \ \left[ \dfrac{ \Gamma^{\text{ads}}_{\text{AI}}(t)} {  A_{\text{drop}}(t) }  \ \diff{ A_{\text{drop}} }{t} \ - \ \diff{  \Gamma^{\text{ads}}_{\text{AI}} }{t}   \right] \\ 
+  \left.  \varepsilon_{\text{AqP}}  \ \eta_{\text{pore}}  \  A_{\Pi}   \ A_{\text{drop}}(t) \ D_{\text{AI}} \ \pder[{  c_{\text{AI}} }]{x} \right\vert_{x=0}  ,  \label{BVCaidrop_3}
\end{multline}
\begin{equation}
\diff{}{t}  \left( V^{\text{drop}}_{\scriptscriptstyle \text{H$_2$O}}(t) \  c_{\text{ADJ}}(t)  \right)  = \dfrac {\xoverline{k}_6 \ A_{\text{drop,0}}}  {\Gamma_S \ A_{\text{drop}}(t)   }   \ \left[  \dfrac{ \Gamma^{\text{ads}}_{\text{ADJ}}(0,t)} {  A_{\text{drop}}(t) }  \ \diff{ A_{\text{drop}} }{t} \ - \ \diff{  \Gamma^{\text{ads}}_{\text{ADJ}} }{t}   \right] ,
   \label{BVCADJdrop_3}
\end{equation}
where $\xoverline{k}_5$ is the kinetic rate constant for the ionic AI and $\xoverline{k}_6$ is the kinetic rate constant for the adjuvant in mol. In Equation (\ref{BVCaidrop_3}), the droplet becomes more concentrated due to evaporation, governed by the volume of water in the drop, $V^{\text{drop}}_{\scriptscriptstyle \text{H$_2$O}}$, and the first term is formulated to incorporate the changing droplet area, $A_{\text{drop}}$. The first term in Equation (\ref{BVCaidrop_3}) is found by substituting Equation (\ref{eq:comp2}) into $\xoverline{k}_5 \ \text{d}\Omega_{\text{AI}} / \text{d}t$ and simplifying. The second term in Equation (\ref{BVCaidrop_3}) describes how ionic AI is transported into the cuticle via diffusion, based on the porosity of the aqueous pores, $\varepsilon_{\text{AqP}}$, the density of aqueous pores in cuticle, $\eta_{\text{pore}}$, the control volume area, $A_\Pi$, the changing area of the droplet, $ A_{\text{drop}}$, the diffusion coefficient for ionic AI, $D_{\text{AI}}$, and the flux of the ionic AI at the droplet boundary. Equation (\ref{BVCADJdrop_3}) does not include the flux term for the concentration of adjuvant, $c_{\text{ADJ}}$, transported into the cuticle, as the adsorbed adjuvant, $\Gamma^{\text{ads}}_{\text{ADJ}}$, is transported into the cuticle in Equation (\ref{eq:ADJads}).

The $\xoverline{k}_5 $ and $\xoverline{k}_6$ rate constants in Equation (\ref{BVCaidrop_3})--(\ref{BVCADJdrop_3}) are as follows:
\begin{align}
\xoverline{k}_5  &= k_5 \ c^{\text{drop}}_{\text{AI,0}}, \label{eq:k5} \\
\xoverline{k}_6 &= k_6 \ c^{\text{drop}}_{\text{ADJ,0}}, \label{eq:k6}
\end{align}
where $k_5$ and $k_6$ are the kinetic rate constants in m$^3$ and $c^{\text{drop}}_{\text{AI,0}}$ and $c^{\text{drop}}_{\text{ADJ,0}}$ are the initial concentration of ionic AI and adjuvant in the drop in mol$/$m$^3$. The parameters $k_5$ and $k_6$ were found to be a function of the initial concentrations, as changing the initial concentration alters the initial droplet contract angle $\theta_0$ (see \citet{Tredenick2018}). 

In summary, the novel Adaptive Competitive Langmuir formulation consists of Equation (\ref{eq:AIads}),(\ref{eq:k6}). The two points of difference from the well-known competitive Langmuir model and the adaptive model described here, are that ionic AI desorbs adjuvant (and vice-versa) and the droplet contact area changes in time, causing some of the adsorbed ionic AI and adjuvant to be left behind by the moving droplet radius and are not available for diffusion. The Adaptive Competitive Langmuir model assumes the following: ionic AI and lipophilic adjuvant adsorb as a monolayer, that the process is competitive, all sites are equivalent, there are no interactions between the adsorbed ionic AI and adjuvant molecules on the cuticle surface and the adsorption--desorption process is not in equilibrium. As this model assumes a competitive process, when one compound is adsorbing and desorbing, then the other must also, so $k_5$ or $k_6$ can only be zero or non-zero pairwise.

\subsection{Numerical Solution Procedure}
The model as described in Equation (\ref{ai3})--(\ref{eq:k6}), is solved numerically. This is achieved using a finite volume method and discretising the model's partial differential equations with second order central differences to approximate the spatial derivatives. The resulting system of ordinary differential equations is solved using `ode15i' \citep{Shampine2002} within MATLAB\textsuperscript{\textregistered} \citep{MATLAB:2016}. The fitted parameter $F_s$ in Table \ref{Parameters_Pr3} was found by using the value in \citet{Tredenick2018}, then the parameter was adjusted by focusing on penetration for only the initial applied mass of 25~$\upmu$g (5~g$/$L) of calcium using trial-and-error. \mbox{The other} fitted parameters in Table \ref{Parameters_Pr3} were found using trial-and-error for only the initial applied mass of 25~$\upmu$g of calcium. The experimental data \citep{Kraemer20092} includes penetration of five different initial applied masses of Ca of 5, 25, 50, 75, and 150~$\upmu$g. All the fitted parameters were then kept constant and used to solve the penetration of the other four initial applied masses.

\section{Results and Discussion} \label{resluts_Pr3}

\begin{figure} [H]
	\centering
	\includegraphics[width=0.75 \textwidth,keepaspectratio=true]{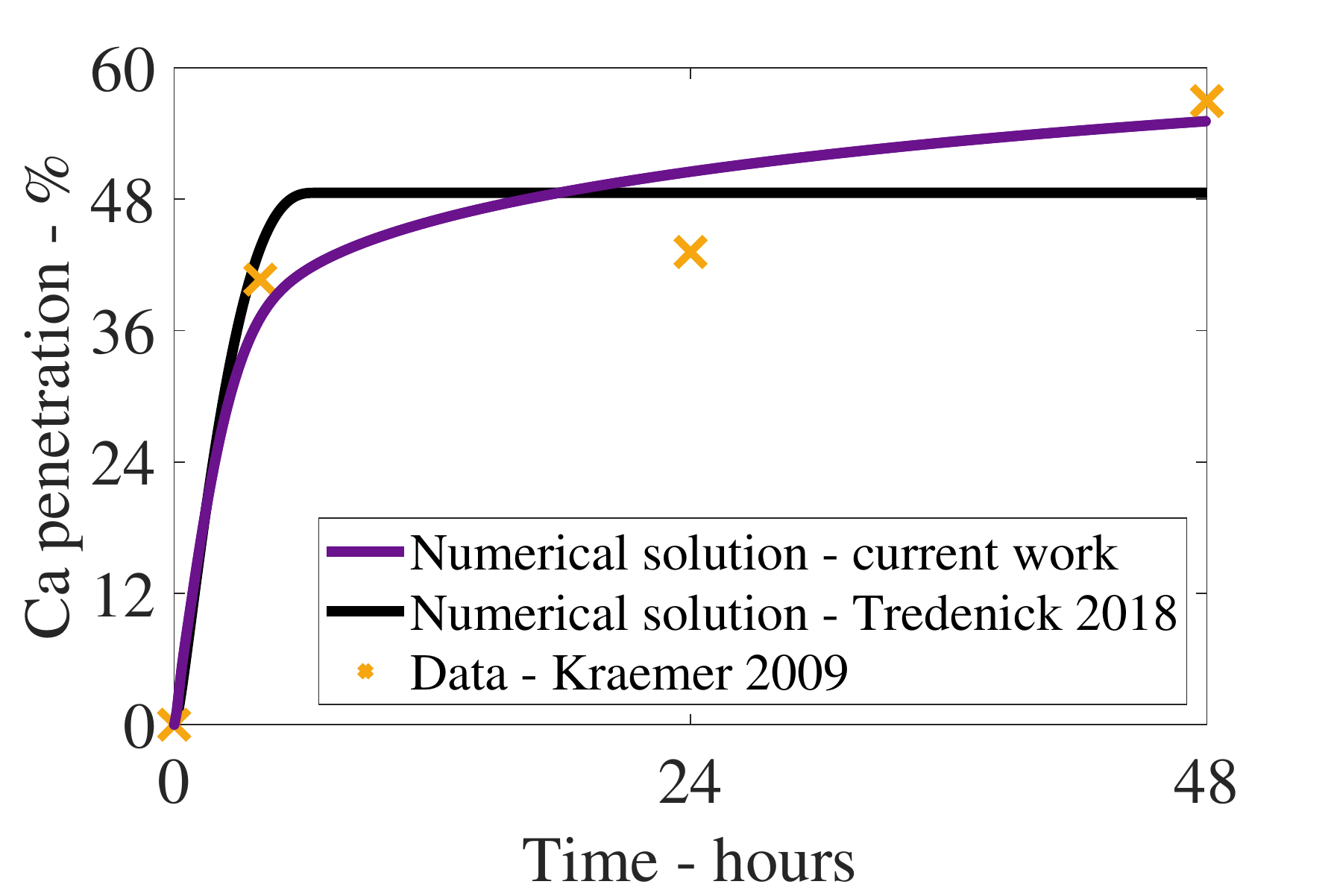}
	\caption[Comparison of penetration.]{Numerical solution results as percentage penetration with time of the model (purple line) compared to experimental data (orange crosses) for \ce{CaCl2} (applied at 5~g$/$L or 25~$\upmu$g) with RSO~5 (1~g$/$L) and the numerical solution of the model in \citet{Tredenick2018} (black line). Parameters are described in Table \ref{Parameters_Pr3}.}
	\label{fig:goingup_latetimes}
\end{figure}

The plant cuticle model, as described in Equation (\ref{ai3})--(\ref{eq:k6}), is solved numerically, with parameters described in Table \ref{Parameters_Pr3}. In Figure \ref{fig:goingup_latetimes}, we can see the numerical solution results as percentage of ionic AI penetration with time (purple line) compared to experimental data (orange crosses---\citep{Kraemer20092}) for \ce{CaCl2} at 5~g$/$L (25~$\upmu$g) with RSO~5 (1~g$/$L). The results of the model compare well to the experimental data, especially at late times, between 4 and 48 h. Penetration increases rapidly over the first 4 h, then begins to level out and the penetration can then continue to increase between 4 and 48 h by $15\%$, due to ionic AI adsorption with desorption. Penetration at 48 h has not yet reached equilibrium and can continue to increase at much later times, beyond 48 h.

In Figure \ref{fig:goingup_latetimes}, the fit of the model in the current work to the experimental data is reasonable; however, the fit is reduced at 24 h. We surmise this could be due to the \citet{Kraemer20092} experimental setup, where each data point represents 10 repeats. Other authors \citep{Schonherr2001ca,Schonherr2001potassium,Schonherr2000} have noted they conducted 50--100 repeats as variability among individual cuticles can be large \citep{Baur1997Long}.

\begin{figure} [H]
	\hspace{-1.5 cm}
	\includegraphics[width=1.2 \textwidth,keepaspectratio=true]{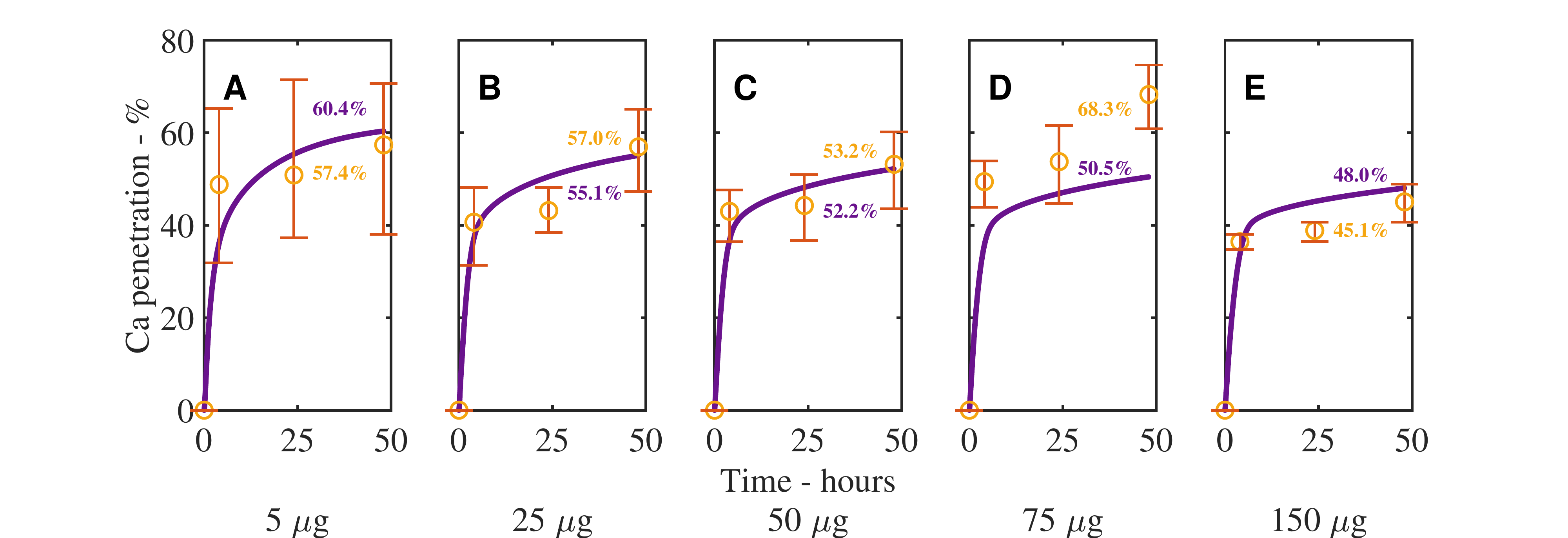}
	\caption[Validation]{Numerical solution (purple line) of the model compared to experimental data (orange circles) for \ce{CaCl2} with RSO~5, using an initial applied mass of \ce{Ca} of 5~$\upmu$g (\textbf{A}), 25~$\upmu$g (\textbf{B}), 50~$\upmu$g (\textbf{C}), 75~$\upmu$g (\textbf{D}), 150~$\upmu$g (\textbf{E}) with RSO~5 (1~g$/$L) over 48 h. Percentage Ca penetration is compared to time in hours, calculated according to Equation (\ref{eq:uptakepercentage3}). The numerical solution can be seen as the continuous purple line and the experimental data as orange circles with error bars. The final percent Ca penetration is shown on each subfigure at 48 h.} 
	\label{fig:velid_Pr3}
\end{figure}

\begin{figure} [H]
	\hspace{-1cm}
	\includegraphics[width=1.1 \textwidth,keepaspectratio=true]{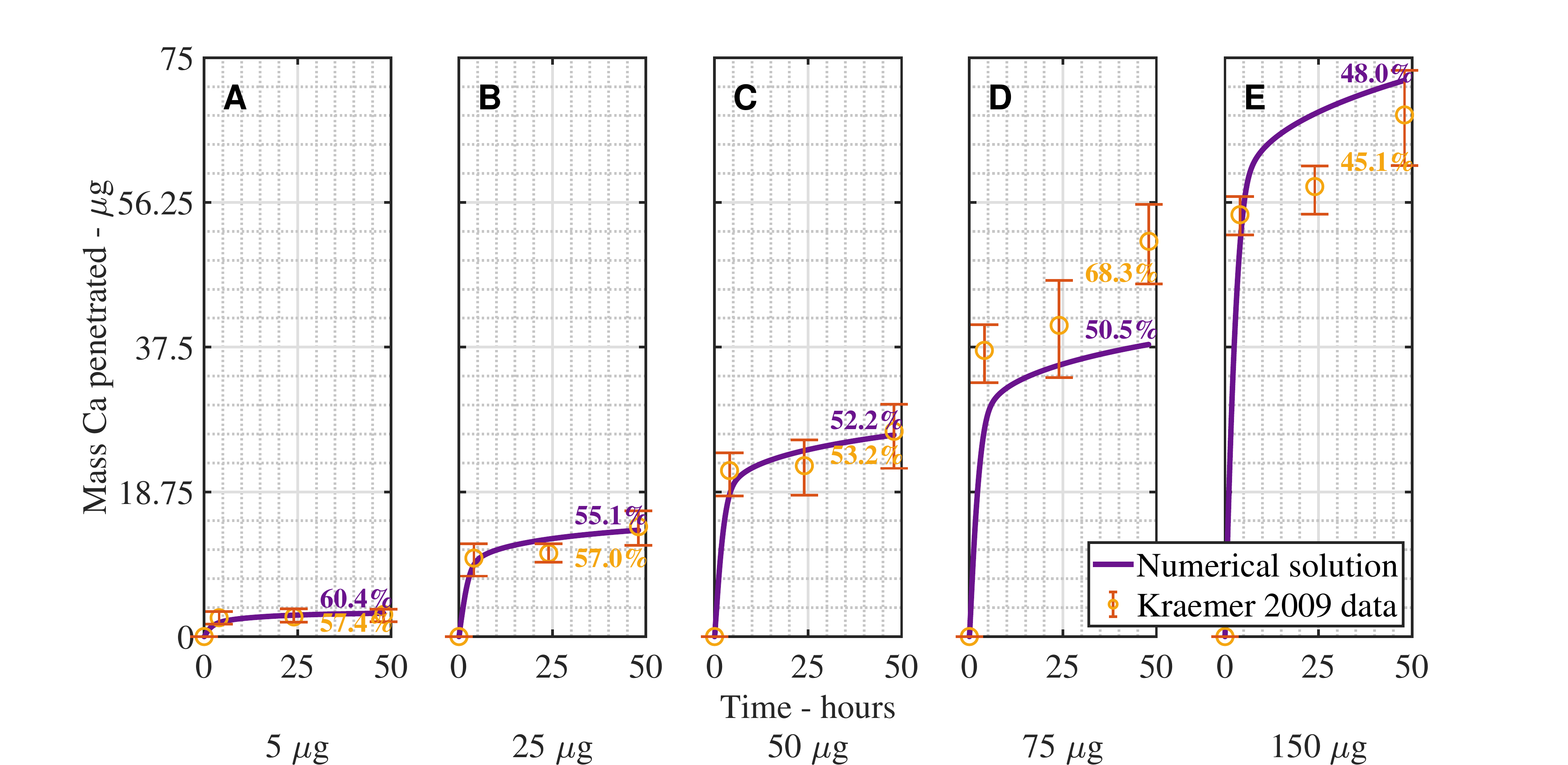}
	\caption[Validation]{{Numerical solution (purple line) of the model compared to experimental data (orange circles) for \ce{CaCl2} with RSO~5, using an initial applied mass of \ce{Ca} of 5~$\upmu$g (\textbf{A}), 25~$\upmu$g (\textbf{B}), 50~$\upmu$g (\textbf{C}), 75~$\upmu$g (\textbf{D}), 150~$\upmu$g (\textbf{E}) with RSO~5 (1~g$/$L) over 48 h. The mass of \ce{Ca} penetrated in $\upmu$g is compared to time in hours, calculated according to Equation (\ref{eq:penetration3}). The numerical solution can be seen as the continuous purple line and the experimental data as orange circles with error bars. The final percent Ca penetration is shown on each subfigure at 48 h.}} 
	\label{fig:velidmass_Pr3}
\end{figure}

In Figure \ref{fig:velid_Pr3},\ref{fig:velidmass_Pr3}, we can see the validation results of the model, compared with experimental data for \ce{CaCl2} with RSO~5, across 5 initial applied concentrations of \ce{CaCl2}. Percentage Ca penetration is compared to time in hours in Figure \ref{fig:velid_Pr3} and as a mass in $\upmu$g in Figure \ref{fig:velidmass_Pr3}. The parameters, as described in Table \ref{Parameters_Pr3}, are the same across all 5 graphs, except the initial applied concentration of \ce{CaCl2}; 5~$\upmu$g (A), 25~$\upmu$g (B), 50~$\upmu$g (C), 75~$\upmu$g (D), 150~$\upmu$g (E). Figure \ref{fig:velid_Pr3} is calculated according to Equation (\ref{eq:uptakepercentage3}) and \mbox{Figure \ref{fig:velidmass_Pr3}} according to Equation (\ref{eq:penetration3}). We have included these two figures as Figure \ref{fig:velidmass_Pr3} shows that increasing the initial concentration of ionic AI increases the mass penetration, while Figure \ref{fig:velid_Pr3} shows the penetration trend clearly.

The results in Figure \ref{fig:velid_Pr3},\ref{fig:velidmass_Pr3} show a reasonable fit across all five plots, considering the complex mechanisms involved, producing an R-squared value of 97.95$\%$, including the error bars. This indicates the model can predict changes in the initial applied concentration. In the figure, the desired trend of increased penetration at late times, after penetration has begun to level out, can be seen in all five graphs.

In Figure \ref{fig:goingup_latetimes}--\ref{fig:velidmass_Pr3}, we can see that the penetration timescale of the current work is significantly extended to more than 48 h. This extended timescale is caused by two mechanisms. The first mechanism is the desorption of ionic AI. At late times, ionic AI is continuously released back into the solution. The second mechanism is the hygroscopic water absorption. As the concentration of ionic AI increases in the drop, due to desorption, hygroscopic water absorption is increased. This in turn extends the evaporation timescale and the penetration timescale. Both the ionic AI desorption and the hygroscopic water absorption promote penetration at late times and both mechanisms must be included in a plant cuticle model for extended penetration to occur at late times.

The increase in penetration in Figure \ref{fig:goingup_latetimes} in the current work, between 4 and 48 h, confirms the addition of the desorption mechanism may be one way to achieve increased penetration at late times. It has been well established that surfactants including RSO increase the penetration of ionic AI, especially at late times between 24 and 100 h, to a significant degree \cite{Kraemer20092,Gauvrit2007,Coret1993}. Earlier work by  \cite{Tredenick2018}, which focused on penetration and droplet evaporation, found that penetration ceased after 9 h, as shown by the black line in Figure \ref{fig:goingup_latetimes}. Penetration ceased after 9 h as the ionic AI was either adsorbed on the cuticle surface to a significant degree or transported through the cuticle. Desorption was not included in the previous work. The current work produces a penetration profile that is closer to the experimental data.

In future works, the model presented here can be adapted to consider a range of constant temperatures. Equation (\ref{diffh2o3})--(\ref{diffADJ3}) for the diffusivities for adjuvant and water in the lipophilic pathway could be modified to be a function of temperature by utilising an Arrhenius equation \citep{laidler1984development}. It has been noted that for lipophilic compounds, when the temperature increases, voids appear and disappear more frequently, which leads to diffusion rates greatly increasing \citep{Schonherr2006,Buchholz2006}. Likewise, an Arrhenius equation could be used to formulate the $k_i$ terms if temperature varied. For lipophilic compounds, it has been found that desorption hysteresis occurs \citep{Schonherr1988} at the inner water bath boundary (in real leaves the boundary condition would represent the apoplast of the epidermal cells), where the adsorbed lipophilic compound may desorb into the bath at a different rate and a portion of compound may be retained within the cuticle. This can be incorporated into the model by utilising a time varying function for $\Gamma^{\text{ads}}_{\text{ADJ}}$ at the bath boundary condition in Equation (\ref{BVCADJbath3}).

{The inner cuticle boundary condition for ionic AI is based on a well stirred water bath, so the concentration of ionic AI is instantaneously removed and is zero in Equation (\ref{BVCaibath3}). In real plants, the inner cuticle surface would be adjacent to the apoplast of the epidermal cells, and transport through the apoplast is thought to be governed by diffusion \cite{sattelmacher2001apoplast,Liu2004}. A flux condition based on diffusion could replace Equation (\ref{BVCaibath3}) for the ionic AI, and a similar equation for the lipophilic adjuvant in Equation~(\ref{BVCADJbath3}).}

{In this work, we have added the mechanism, where the lipophilic adjuvant causes the ionic AI to desorb from the cuticle surface, and this increases penetration at late times. Without lipophilic adjuvant, desorption of ionic AI is not included, and the model can be applied when no adjuvant is included in the formulation by setting the initial applied concentration of adjuvant in Equation (\ref{IVPADJdrop3}) to~zero.}

The fitted model parameters are shown in Table \ref{Parameters_Pr3}. Some of these parameters have physiological significance, such as the tortuosity or fractal scaling dimension, $F_s$. A low value for $F_s$ of 1.15 was found with the fitting exercise. A low value for $F_s$ indicates the aqueous pores in tomato fruit cuticles are in the lower range of tortuosity, tending to straight. Penetration is quite rapid, even without adjuvant, so this value is logical. Fitting $F_s$ is reasonable as it facilitates the diffusion path length calculation, which currently cannot be found by a physical measurement \citep{Riederer1995thick}. The value for the partial molar volume of the adjuvant, $\bar{v}_{\text{ADJ}}$, was not available in the literature and was found by fitting with trial-and-error. Likewise, the value for the bulk diffusion coefficient of the adjuvant, $D_{\text{ADJ}}^{\text{bulk}}$, was not available in the literature and was found by fitting with trial-and-error. The value for the porosity of the lipophilic pathway, $\varepsilon_{\text{L}}$, was found to be similar to the value for the porosity of the aqueous pathway, $\varepsilon_{\text{AqP}}$. The~values in \citet{Tredenick2018} for the aqueous pore density, $\eta_{\text{pore}}$, and logistic decay evaporation constant, $\chi$, were utilised. We note the model has many fitted parameters. However, the penetration of ionic AI showed little sensitivity to $D_{\text{ADJ}}^{\text{bulk}}$ and $\varepsilon_{\text{L}}$, when fitting with trial-and-error. The~parameters $D_{\text{ADJ}}^{\text{bulk}}$, $\bar{v}_{\text{ADJ}}$ and $\Gamma_{\text{S}}$ may be found with further experimental works, reducing the number of fitted parameters. {The parameters $D_{\text{ADJ}}^{\text{bulk}}$ and $\bar{v}_{\text{ADJ}}$ for RSO~5 can then be used in any material, not just plant cuticles.}

\subsection{Sensitivity Analysis}\label{SA_PR3}

\begin{figure} [H]
	\centering
	\includegraphics[width=1 \textwidth,keepaspectratio=true]{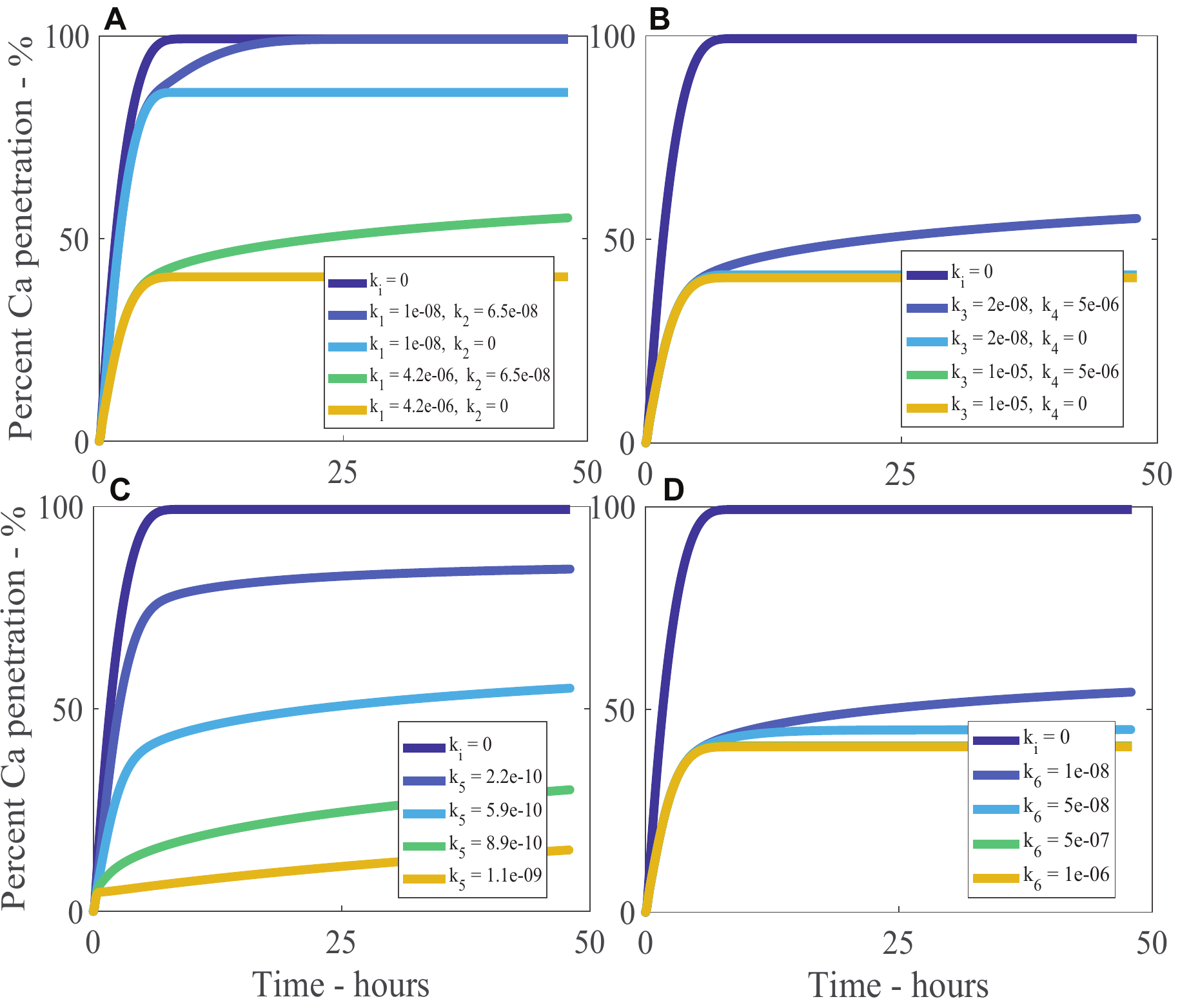}
	\caption{Percentage of calcium (Ca) penetration sensitivity to $k_1$ and $k_2$ (\textbf{A}), $k_3$ and $k_4$ (\textbf{B}), $k_5$ (\textbf{C}) and $k_6$ (\textbf{D}), with parameters described in Table \ref{Parameters_Pr3} at an initial concentration of \ce{CaCl2} of 5~g$/$L and RSO~5 of 1~g$/$L. The kinetic rate constants for the ionic AI are $k_1$ and $k_2$ (\textbf{A}) and $k_5$ (\textbf{C}). The kinetic rate constants for the lipophilic adjuvant are $k_3$ and $k_4$ (\textbf{B}) and $k_6$ (\textbf{D}).}
	\label{k_6_SA} \label{k_5_SA} \label{k_34_SA} 	\label{k_1_2_SA}
\end{figure}

A sensitivity analysis was performed with the results from the model, using parameters in Table \ref{Parameters_Pr3} at an initial concentration of \ce{CaCl2} of 5~g$/$L (or 25~$\upmu$g) with RSO~5 (1~g$/$L). We have investigated the adsorption--desorption parameters, with the method described in \mbox{\citet{Tredenick2017}} and \mbox{\citet{Tredenick2018}}. The results of the sensitivity analysis are seen in Figure \ref{k_1_2_SA}A--D, where dimensionless percentage penetration of ionic AI or percent calcium (Ca) penetration is investigated over a timescale of 48 h. A representative range of $k_i$ values are investigated, where $i$ is between 1 and 6, along with dimensionless penetration.

In Figure \ref{k_1_2_SA}A, we investigate the sensitivity of the models results for percentage penetration with the adsorption and desorption rate constants for ionic AI; $k_1$ and $k_2$. In the figure, the ionic AI adsorption rate constant, $k_1$, controls the percent penetration where the penetration starts to level out and is inversely proportional to penetration. A larger value for $k_1$ indicates more ions are being adsorbed so penetration is reduced. The ionic AI desorption rate constant, $k_2$, is directly proportional to penetration and influences the amount of desorption of ionic AI. If $k_2$ is zero, penetration levels out earlier and no desorption occurs. As $k_2$ increases, more ionic AI is released at later times, which increases penetration and the penetration timescale is extended. Only when no adsorption occurs \mbox{($k_i=0$)} can penetration reach 100$\%$. The sensitivity in Figure \ref{k_1_2_SA}A aligns to a similar adsorption--desorption model in the well-established literature \citep{Cameron1977}.

In Figure \ref{k_34_SA}B, we investigate the sensitivity of the model's results for ionic AI percentage penetration with the adjuvant adsorption and desorption rate constants; $k_3$ and $k_4$. In the figure, the adjuvant adsorption rate constant, $k_3$, has minimal influence over AI penetration and is inversely proportional to penetration. As $k_3$ increases, ionic AI penetration decreases as more adjuvant is adsorbed, which decreases the concentration of adjuvant in the drop and this causes less ionic AI to desorb and penetration ceases earlier. The adjuvant desorption rate constant, $k_4$, has minimal influence over ionic AI penetration and is directly proportional to penetration. As $k_4$ increases, ionic AI penetration increases at late times as more adjuvant is being desorbed, which increases the concentration of adjuvant in the drop and increases ionic AI desorption.

Figure \ref{k_5_SA}C shows the sensitivity of the model to the kinetic rate constant for ionic AI, $k_5$, which controls how influential the adsorption and desorption of ionic AI is, when included into the droplet boundary condition in Equation (\ref{BVCaidrop_3}). Penetration of ionic AI is very sensitive to $k_5$ and penetration is inversely proportional to $k_5$. As the value of $k_5$ increases, both adsorption and desorption increase. As adsorption increases, penetration levels out at a lower value. Increasing $k_5$ also increases the amount of desorption, which can be seen as the gradient for $k_5$ at $\num{1.1e-9}$ is larger than at $\num{2.2e-10}$. A balance needs to be obtained between adsorption and desorption to explain the experimental data~trends.

Figure \ref{k_6_SA}D shows the sensitivity of the model results to the kinetic rate constant for adjuvant, $k_6$, which controls the level of adsorption and desorption of adjuvant. Penetration is inversely proportional to $k_6$. When $k_6$ is small, this reduces both the adsorption and desorption of adjuvant, resulting in the concentration of adjuvant in the droplet increasing. This causes an increased concentration of adjuvant in the droplet solution that increases ionic AI desorption, which increases ionic AI penetration at late times. Conversely, when $k_6$ is large, both the adsorption and desorption of adjuvant are increased and the increase of adjuvant adsorption causes the concentration of adjuvant in the drop to significantly reduce and thus decreases ionic AI desorption. Therefore, a balance needs to be maintained.

The parameters discussed in the sensitivity analysis show appropriate sensitivities within the model and it is clear that a balance needs to be maintained between adsorption and desorption of ionic AI. We note that there are few experimental studies that measure the adsorption and desorption kinetic rate constants for adjuvants. Future experimental works, similar to \citet{Yamada1964}, could be conducted to directly measure the $k_i$ rate constants for RSO~5 and identify the mechanisms that are involved in adjuvant transport through plant leaves and cuticles.

The Adaptive Competitive Langmuir model applies to sets of two substances, which are adsorbing and desorbing in a competitive manner, under a dynamic droplet contact area. The plant cuticle diffusion model can apply to the penetration of most hydrophilic ionic AIs including lipophilic adjuvants in the formulation. It can apply to isolated astomatous plant leaf or fruit cuticle, where the aqueous pores are sufficiently large to allow ionic AI to be transported through the cuticle via Fickian diffusion. The model for lipophilic transport of adjuvant through the cuticle can be applied to both lipophilic adjuvants and lipophilic AIs. Adaptations can be made for hydrophilic adjuvants as transport is thought to be similar to hydrophilic ionic AIs. The model cannot be applied, in its current form, to whole leaf penetration, hydrophilic uncharged compounds, nor compounds, such as Fe chelates \citep{Schonherr2005, Schlegel2006rates}, that dehydrate aqueous pores. Adaptations could be made to this model to account for these compounds and this is the subject of future work.{ We note that if the type of hydrophilic ionic AI, lipophilic adjuvant, or plant species was changed, the model and mechanisms would not change. However, new parameter values would need to be found, and this may present a challenge depending on the extent of the available experimental literature.}

\section{Conclusions}\label{Sec:concl}
We created a novel Adaptive Competitive Langmuir model that allows a dynamic droplet contact area, which influences the amount of adsorption and desorption of molecules. A mechanistic model has, therefore, been developed to simulate diffusion of hydrophilic ionic AIs, including lipophilic adjuvants through plant cuticles. This model makes novel additions to a simple diffusion model by incorporating the important governing mechanisms of the adsorption and desorption of ionic AI and adjuvant at the plant cuticle surface, along with the adsorption and subsequent diffusion of the adsorbed lipophilic adjuvant through the cuticle. The model has been solved numerically, producing results that show reasonable agreement with the experimental data. The~addition of the desorption mechanism at the cuticle surface enables penetration of AI to increase at later times and further explains the experimental data trends. The sensitivity analysis indicated the parameters that controlled the adsorption and desorption had a significant impact on penetration. The results indicate the chemistry within the droplet on the cuticle surface is significant. Overall, the results of the validation and sensitivity analysis imply the model has included many important governing mechanisms simulating ionic AI penetration through plant cuticles.

\vspace{6pt}

\authorcontributions{conceptualisation, E.C.T., T.W.F., W.A.F.; methodology, E.C.T., T.W.F., W.A.F.; computational code creation, E.C.T., T.W.F.; model creation and adaptation, E.C.T., T.W.F.; validation, E.C.T., T.W.F.; formal analysis, E.C.T., T.W.F.; writing---original draft preparation, E.C.T.; writing---review and editing, E.C.T., T.W.F., W.A.F.}

\funding{This research received no external funding.}

\acknowledgments{We thank S.T.P.~Psaltis for his contribution to the partial development of the numerical~scheme.}

\conflictsofinterest{The authors declare no conflict of interest.} 

\appendixtitles{yes} 
\appendix
\section{Additional Droplet Evaporation Equations}\label{Appendix1}
	The following describe the additional droplet evaporation equations and further details can be found in \cite{Tredenick2018}:
	\begin{align*}
	A_{\scriptscriptstyle \text{drop,0}}  &= \pi^{\frac{1}{3}} \left(  3 \ g(\theta_0) \ V_0 \right) ^{\frac{2}{3}},\\
	c_{\text{mass}\%} &= -0.8307 \ \mathrm{e}^{3.618 \ \Phi} + 55.44 \ \mathrm{e}^{-0.612 \ \Phi},\\
	c_{\text{POD}} &= \dfrac{ c_{\text{mass}\%} \  \rho_{\scriptscriptstyle \text{AI}}  }{ 100\% \  M_{\text{w,AI}}},\\
	g(\theta) &= \dfrac{\sin^3(\theta)}  { (1-\cos(\theta))^2 \ (2+\cos(\theta))},\\
	m_\infty &= 0.307 \ \mathrm{e}^{2.763 \ \Phi} +  1.218 \times 10^{-9} \ \mathrm{e}^{24 \ \Phi},\\
	r_{\text{drop,0}} &= \left( \dfrac{ 3 \ g(\theta_0) \ V_0} { \pi} \right) ^{\frac{1}{3}},\\
	V_{\text{Del}} &=  \dfrac{ m_\infty \ M_{\text{w,AI}} \  c_{\text{AI}}(0,t)  \  V^{\text{drop}}_{\scriptscriptstyle \text{H$_2$O}}  }{ \rho_{\scriptscriptstyle \text{H$_2$O}}   },   \\
	\xoverline{\chi} &= \chi \ c^{\text{drop} \ 2}_{\text{AI,0}},\\
	\Lambda &= \dfrac{ D_{\text{evap}} \ \psi } { \rho_{\scriptscriptstyle \text{H$_2$O}} },\\ 
	\psi  &= \dfrac{ M_{\scriptscriptstyle \text{w,H$_2$O}} \ P_{\text{v}} (1-H) }{ R \ T },
	\end{align*}
	\begin{align*}
	f(\theta) &= \tan\left(  \dfrac{\theta}{2} \right) + 8 \int\limits_0^{\infty} \cosh^2(\theta u) \ \text{csch}(2  \pi  u) \ \tanh \left[ \left(  \pi - \theta   \right) u \right] \ du, \\
	t_{\text{rec}} & =  \int\limits_{\theta_{\text{rec}}}^{\theta_0} \dfrac{r_{\text{drop,0}}^2}{ \Lambda \ \left(  1 + \cos(\theta) \right)^2 \ f(\theta)  }  \ d\theta. 
	\end{align*}
\pagebreak
\reftitle{References}

\end{document}